\documentclass[twocolumn,showpacs,preprintnumbers,amsmath,amssymb,nofootinbib]{revtex4}

\usepackage{graphicx}
\usepackage{dcolumn}
\usepackage{bm}
\usepackage{color}

\newcommand{\eq}[1]{Eq.~(\ref{#1})}

\newcommand{\bdtau}{\ensuremath{B\to D\tau\nu\;}}
\newcommand{\bdstau}{\ensuremath{B\to D^*\tau\nu\;}}
\newcommand{\btau}{\ensuremath{B\to \tau\nu\;}}
\newcommand{\rd}{\ensuremath{{\cal R}(D)}\;}
\newcommand{\rds}{\ensuremath{{\cal R}(D^*)}\;}

\begin{document}

\title{Explaining \bdtau, \bdstau and \btau in a two Higgs doublet model of type III}
%
\author{Andreas Crivellin, Christoph Greub and Ahmet Kokulu}
\affiliation{Albert Einstein Center for Fundamental Physics,\\ Institute for Theoretical Physics,\\
              University of Bern,\\ CH-3012 Bern, Switzerland}
%
\begin{abstract}
Recently, the BABAR collaboration reported first evidence for new physics in \bdtau and \bdstau. Combining both processes, the significance is $3.4\, \sigma$. This result cannot be explained in a two Higgs doublet model of type II. Furthermore, the CKMfitter Group finds a $2.9\, \sigma$ discrepancy between the Standard Model prediction for Br[\ensuremath{B\to \tau\nu}] \;(using $V_{ub}$ from a global fit to the unitary triangle) and the measurements of the $B$ factories. Altogether, these measurements are strong indications for physics beyond the Standard Model in $B$-meson decays to taus.

We show that in a two Higgs doublet model of type III it is possible to simultaneously explain \bdtau and \bdstau using a single free parameter $\epsilon^u_{32}$. Also, Br[\ensuremath{B\to \tau\nu}] \;can be brought into agreement with experiment using $\epsilon^u_{31}$. Furthermore, for Higgs ($A^0, H^0, H^\pm$) masses around 500 GeV, as preferred by recent CMS results, all bounds from FCNC processes are satisfied and \bdtau, \bdstau and \btau can be explained without a significant degree of fine tuning.

\end{abstract}
\maketitle

\vspace{-1mm}
\section{\label{sec:level1}Introduction}

In addition to the direct searches for new physics (performed at very high energies) at the LHC, low-energy precision flavour observables provide a complementary window to physics beyond the Standard Model (SM). Tauonic $B$-meson decays are an excellent probe of new physics: they test lepton flavor universality satisfied in the Standard Model (SM) and are sensitive to new particles which couple proportionally to the mass of the involved particles (e.g. Higgs bosons) due to the heavy $\tau$ lepton involved. The single decay modes still suffer from large hadronic uncertainties related to the form factors and from the uncertainties of the CKM elements. However, in normalizing the $\tau$ decay mode to the corresponding decay with light leptons in the final state, these uncertainties are reduced and the sensitivity to new physics is significantly improved.

Recently, the BABAR Collaboration performed an analysis of the semileptonic $B$ decays \bdtau and \bdstau using the full available data set \cite{BaBar:2012xj}. They find for the ratios
\begin{equation}
{\cal R}(D^{(*)})\,=\,{\cal B}(B\to D^{(*)} \tau \nu)/{\cal B}(B\to D^{(*)} \ell \nu)\,,
\end{equation}
the following results:
\begin{eqnarray}
{\cal R}(D)\,=\,0.440\pm0.058\pm0.042  \,,\\
{\cal R}(D^*)\,=\,0.332\pm0.024\pm0.018\,.
\end{eqnarray}
Here the first error is statistical and the second one is systematic. Comparing these measurements to the SM predictions
\begin{eqnarray}
{\cal R}_{\rm SM}(D)\,=\,0.297\pm0.017 \,, \\
{\cal R}_{\rm SM}(D^*) \,=\,0.252\pm0.003 \,,
\end{eqnarray}
we see that there is a discrepancy of 2.2\,$\sigma$ for \rd and 2.7\,$\sigma$ for \rds. For these theory predictions we again used the updated results of \cite{BaBar:2012xj}, which rely on the calculations of Refs.~\cite{Kamenik:2008tj,Fajfer:2012vx} based on the previous results of Refs.~\cite{Korner:1987kd,Korner:1989ve,Korner:1989qb,Heiliger:1989yp,Pham:1992fr}. Both processes exceed the SM prediction, and combining them gives a $3.4\, \sigma$ deviation from the SM~\cite{BaBar:2012xj}, which constitutes the first evidence for new physics in semileptonic B decays to tau leptons.

This evidence for new physics in $B$-meson decays to taus is further supported by the measurement of \btau by BABAR \cite{Aubert:2008zzb} and BELLE \cite{Hara:2010dk}. Averaging both measurements, one obtains the branching ratio \cite{Asner:2010qj}
\begin{equation}
{\cal B}[B\to \tau\nu]=(1.67\pm0.3)\times 10^{-4}\,.
\end{equation}
This also disagrees with the SM prediction by $2.9\, \sigma$ \cite{Charles:2004jd} or $2.5\, \sigma$ \cite{Bona:2009cj}, using the global fit of the CKM matrix performed by CKMfitter or UTfit, respectively.

Thus, combining \rd, \rds and \btau, we have rather solid evidence for violation of lepton flavor universality. Assuming that these deviations from the SM are not statistical fluctuations or underestimated theoretical or systematic errors, it is interesting to ask which model of new physics can explain the measured values. Since these processes are all tree-level decays in the SM, it is difficult to explain these deviations with a model of new physics (NP), since one in general also needs a tree-level exchange of a new particle in order to get sizable effects. This then generates the difficulty to explain the absence of NP effects in other observables.

A widely studied possibility is the introduction of a charged scalar particle which couples proportionally to the masses of the fermions involved in the interaction: a charged Higgs boson. Such a charged Higgs boson is introduced in the MSSM or in general in any two Higgs doublet model (2HDM), and affects \btau~\cite{Hou:1992sy,Akeroyd:2003zr}, \bdtau and \bdstau~\cite{Tanaka:1994ay,Miki:2002nz,Nierste:2008qe}. This is a reasonable model: because the Higgs couples only significantly to the tau, it can explain the absence of NP effects in $B$ decays to light leptons and gives rise to lepton flavor universality violation.

In a 2HDM of type II (like the MSSM\footnote{At the loop-level non-holomorphic couplings are induced, but for constructive interference they have to exceed the tree-level Yukawa coupling which is very difficult.}), one Higgs doublet couples to down quarks and charged leptons, while the other one gives masses to the up quarks. Then the only free additional parameters are $\tan\beta=v_u/v_d$ (the ratio of the two vacuum expectation values) and the charged Higgs mass $m_{H^\pm}$ (the heavy CP even Higgs mass $m_{H^0}$ and the CP odd Higgs mass $m_{A^0}$ can be expressed in terms of the charged Higgs mass and differ only by electroweak corrections). In this setup the charged Higgs contribution to \btau interferes necessarily destructively with the SM \cite{Hou:1992sy}. Thus, an enhancement of $\cal B$[$B\to \tau\nu$] is only possible if the absolute value of the charged Higgs contribution is bigger than two times the SM one, which is in conflict with \bdtau. Furthermore, a 2HDM of type II cannot explain \rd and \rds simultaneously \cite{BaBar:2012xj}.

Another possibility to explain \btau is the introduction of a right-handed $W$-coupling \cite{Crivellin:2009sd} or new physics in $B$ mixing \cite{Lenz:2012az} (meaning that the actual value of $V_{ub}$ is bigger than the one extracted from the global fit). Anyway, neither possibilities can help to explain the deviation from the SM in \rd and \rds.

Thus, we need another model to explain \rd and \rds. Our choice in this article is a 2HDM of type III (where both Higgs doublets couple to up quarks and down quarks as well) with MSSM-like Higgs potential. Since a 2HDM of type III with minimal flavor violation (MFV) can only explain \btau in some fine-tuned regions of parameter space \cite{Blankenburg:2011ca} and cannot explain \rd and \rds simultaneously, we consider a more generic flavor structure with flavor violation in the up sector. As we will see, this model is capable to explain \btau, \rd and \rds without fine tuning.

\section{Effective Field Theory}

\begin{figure*}[t]
\centering
\includegraphics[width=0.3\textwidth]{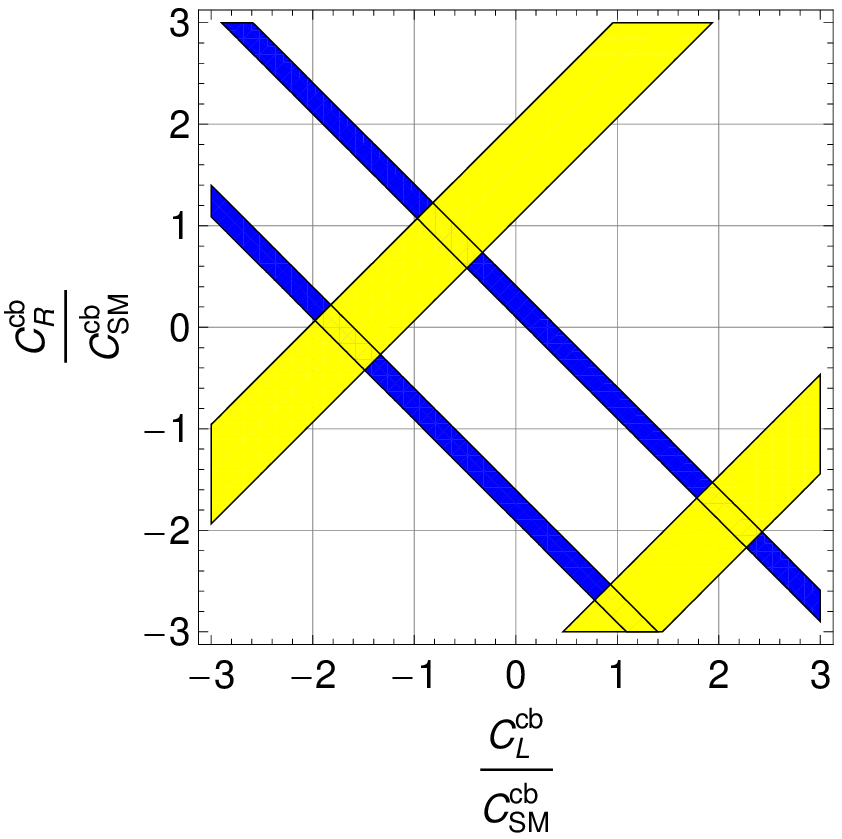}
\includegraphics[width=0.3\textwidth]{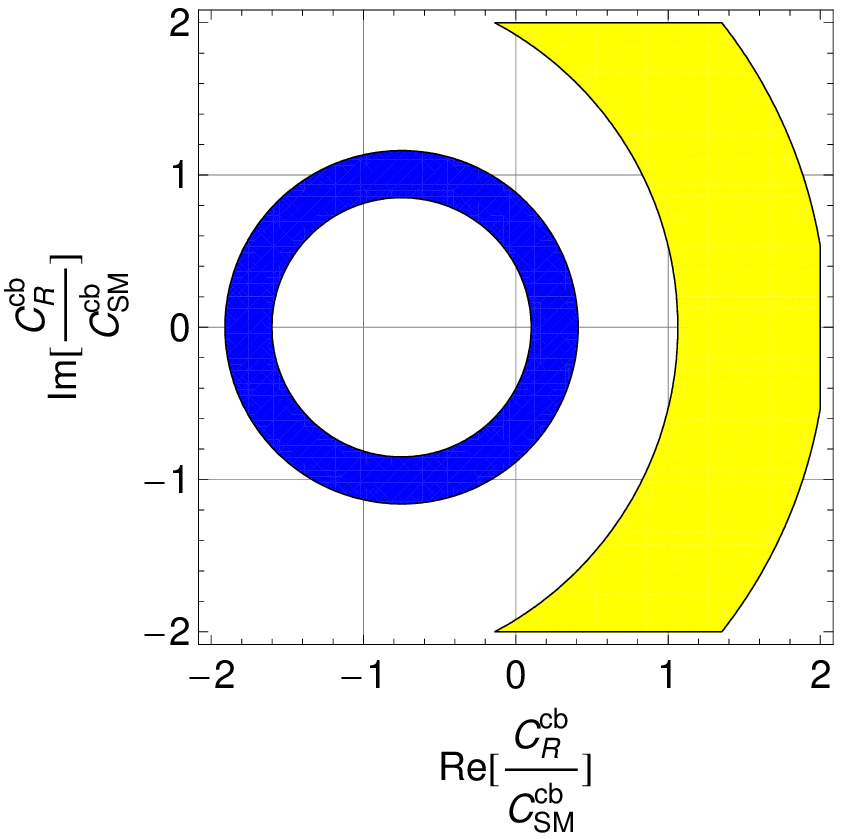}
\includegraphics[width=0.3\textwidth]{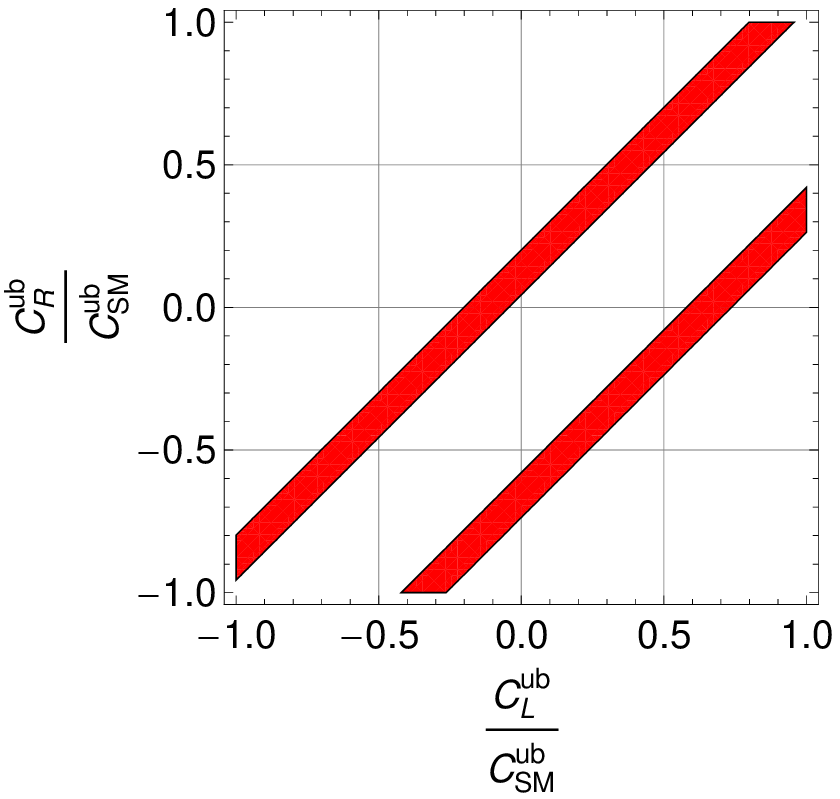}
\caption{Left and middle: Allowed $1\, \sigma$ regions from \rd (blue) and \rds (yellow), adding the experimental uncertainty and theoretical uncertainty linear. Left: Constraints in the $C_L^{cb}/C^{cb}_{\rm SM}$--$C_R^{cb}/C^{cb}_{\rm SM}$ plane for real values of $C_L^{cb}/C^{cb}_{\rm SM}$ and $C_R^{cb}/C^{cb}_{\rm SM}$. Middle: $C_R^{cb}$ complex for $C_L^{cb}$=0. Right: Allowed $1\, \sigma$ regions from \btau in the $C_L^{ub}/C^{ub}_{\rm SM}$--$C_R^{ub}/C^{ub}_{\rm SM}$ plane for real values of $C_L^{ub}/C^{ub}_{\rm SM}$ and $C_R^{ub}/C^{ub}_{\rm SM}$. All Wilson coefficients are understood to be at the scale $m_b$.
\label{EFT}}
\end{figure*}
  
Since the NP we are interested in must be far above the scale of the $B$ meson, we can integrate out the heavy degrees of freedom (including the SM $W$ boson). The SM contribution and the NP contribution are then contained within the effective Hamiltonian
\begin{equation}
\renewcommand{\arraystretch}{1.8}
\begin{array}{l}
 {\cal H}_{\rm eff}=  C^{qb}_{\rm SM} O_{{\rm SM}}^{qb} +   C_{R}^{qb} O_{R}^{qb} + C_{L}^{qb} O_{L}^{qb}\,,
\end{array}
\label{Heff}
\end{equation}
with (for massless neutrinos)
\begin{equation}
\renewcommand{\arraystretch}{1.8}
\begin{array}{l}
   O_{{\rm SM}}^{qb}  = \bar q\gamma _{\mu } P_L b \; \bar\tau \gamma_{\mu } P_L \nu_{\tau}\,, \\ 
   O_{R}^{qb}  = \bar q P_R b \; \bar\tau P_L \nu_{\tau}\,, \\
O_{L}^{qb}  = \bar q P_L b \; \bar\tau P_L \nu_{\tau}\,.   \\
\end{array}
\label{Oeff}
\end{equation}
In \eq{Heff} and \eq{Oeff} $q=u$ for \btau and $q=c$ for \bdtau and \bdstau. The SM Wilson coefficient is given by $ C_{{\rm SM}}^{qb} = { 4 G_{F}} \; V^{}_{qb}/{\sqrt{2}}$. 
The corresponding Wilson coefficients $C_{R}^{qb}$ and $C_{L}^{qb}$ (given at the $B$ meson scale), {which parametrize the effect of NP}, affect our three physical observables in the following way \cite{Akeroyd:2003zr,Fajfer:2012vx,Sakaki:2012ft}:
\begin{widetext}
\begin{eqnarray}
{\cal R}(D) = {\cal R}_{\rm SM}(D) \left(1 + 1.5 \Re\left[\frac{C_{R}^{cb}+C_L^{cb}}{C_{SM}^{cb}}\right]+1.0 \left|\frac{C_{R}^{cb}+C_L^{cb}}{C_{SM}^{cb}}\right|^2  \right)\,,\\
{\cal R}(D^*) = {\cal R}_{\rm SM}(D^*) \left(1 + 0.12 \Re \left[\frac{C_{R}^{cb}-C_L^{cb}}{C_{SM}^{cb}}\right]+0.05 \left|\frac{C_{R}^{cb}-C_L^{cb}}{C_{SM}^{cb}}\right|^2  \right)\,,\\
\mathcal B [B\to\tau\nu] = \dfrac{G_F^2|V_{ub}|^2}{8\pi} m_\tau^2 f_B^2 m_B \left(1-\dfrac{m_\tau^2}{m_B^2}\right)^2 \tau_B    
\times \left| 1+ \frac{m_B^{2}}{\overline{m}_{b}m_{\tau}} \frac{(C_{R}^{ub}-C_L^{ub})}{C_{SM}^{ub}}   \right|^2\,.
\end{eqnarray}
\end{widetext}
Let us consider first \bdtau and \bdstau, where the ratios \rd and \rds are affected by the two Wilson coefficients $C_{R}^{cb}$ and $C_{L}^{cb}$. For our analysis we add the experimental errors in quadrature and the theoretical uncertainty linear on top of this. From the left plot in Fig.~\ref{EFT}, we see that both \rd and \rds can be brought into agreement with the experimental values within the $1\, \sigma$ error by $C_L^{cb}$ only. Note that $C_R^{cb}$ is not capable of achieving this without a simultaneous contribution from $C_L^{cb}$. Since (neglecting small mass ratios) only $C_R^{cb}$ is generated in a 2HDM of type II or in a 2HDM of type III with MFV \cite{MFV} (neglecting small quark mass ratios), these models cannot explain \rd and \rds simultaneously. This is still true if we allow for complex values of $C_R^{cb}$, as we can see from the middle plot in Fig.~\ref{EFT}. Note that the Wilson coefficients in the plots are given at the scale $m_b$.

On the other hand, \btau can be explained either with $C_R^{ub}$ or with $C_L^{ub}$ (or with a combination of both of them). However, as we will see in the next section, in the context of the 2HDM of type III, $C_L^{ub}$ is the more natural choice.

\section{Two Higgs doublet model of type III}

\begin{figure}[t]
\centering
\includegraphics[width=0.4\textwidth]{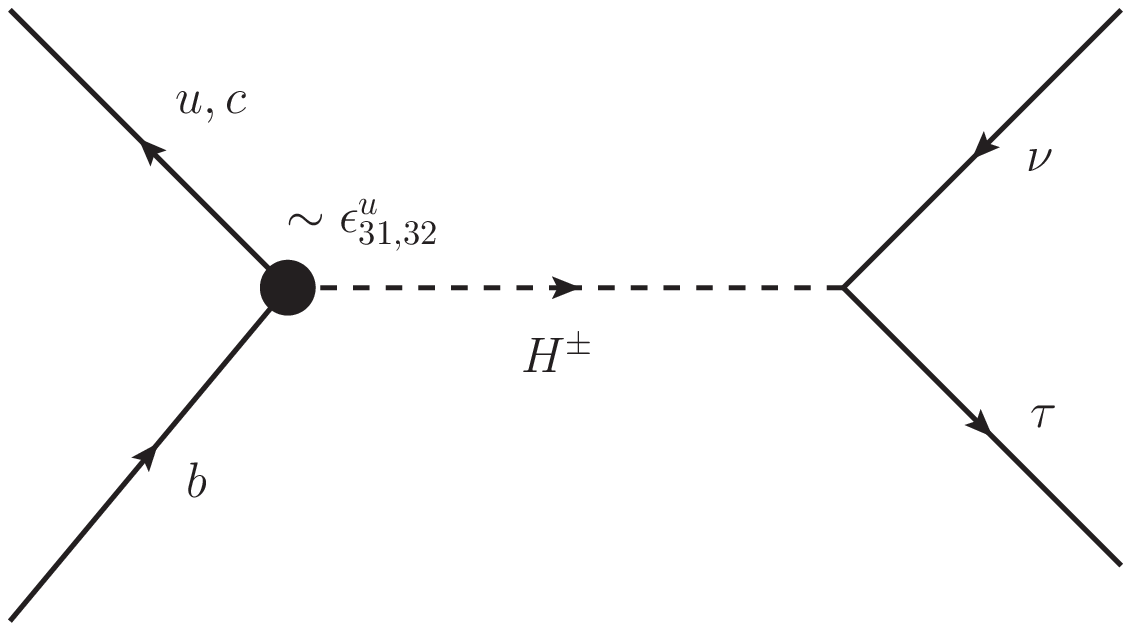}
\caption{Feynman diagram with a charged Higgs contributing to \btau and $B\to D^{(*)}\tau\nu$. The dot represents the flavor-violating interaction containing the 2HDM of type III parameters $\epsilon^u_{31}$ and $\epsilon^u_{32}$, which affect \btau and $B\to D^{(*)}\tau\nu$, respectively.
\label{feynman-diagram}}
\end{figure}

The SM contains only one scalar isospin doublet, the Higgs doublet. After electroweak symmetry breaking, this gives masses to up quarks, down quarks and charged leptons. The charged component of this doublet becomes the longitudinal component of the $W$ boson, and thus we have only one physical neutral Higgs particle. In a 2HDM we introduce a second Higgs doublet and obtain four additional physical Higgs particles (in the case of a CP conserving Higgs potential): the neutral CP-even Higgs $H$, a neutral CP-odd Higgs $A$ and the two charged Higgses $H^{\pm}$.

Two Higgs doublet models have been studied for many years with focus on the type II models \cite{Miki:2002nz,WahabElKaffas:2007xd,Deschamps:2009rh} or type III models with MFV \cite{MFV,Buras:2010mh,Blankenburg:2011ca}, and on alignment \cite{Pich:2009sp,Jung:2010ik} or natural flavour conservation \cite{Glashow:1976nt,Buras:2010mh}. As outlined in the introduction, these models cannot explain \rd and \rds simultaneously \cite{BaBar:2012xj} (and for \btau fine tuning is needed); we will study a 2HDM of type III with generic flavour-structure \cite{Cheng:1987rs}, but for simplicity, with MSSM-like Higgs potential \footnote{Flavor-observables in type III models have been considered before \cite{Mahmoudi:2009zx}, but with focus on the flavor-changing elements in the down sector.}.
 
In the 2HDM of type III, we have the Yukawa Lagrangian (see for example \cite{Crivellin:2010er} for details):
\begin{eqnarray}
\mathcal{L}^{eff}_Y &=& \bar{Q}^a_{f\,L} \left[
  Y^{d}_{fi} \epsilon_{ab}H^{b\star}_d\,-\,\epsilon^{d}_{fi} H^{a}_u \right]d_{i\,R}\\
&-&\bar{Q}^a_{f\,L} \left[ Y^{u}_{fi}
 \epsilon_{ab} H^{b\star}_u \,+\, \epsilon^{ u}_{fi} H^{a}_d
  \right]u_{i\,R}\,+\,\rm{H.c}. \,,\nonumber
\end{eqnarray}
where $\epsilon_{ab}$ is the totally antisymmetric tensor, and $\epsilon^q_{ij}$ parametrizes the non-holomorphic corrections which couple up (down) quarks to the down (up) type Higgs doublet. After electroweak symmetry breaking, this Lagrangian gives rise to the following Feynman-rule:
\begin{equation}
i\left({\Gamma_{u_f d_i }^{H^\pm\,LR\,\rm{eff} } }P_R+{\Gamma_{u_f d_i }^{H^\pm\,RL\,\rm{eff} } }P_L\right)\, ,\\
 \label{Higgs-vertex}
\end{equation}
with
\begin{eqnarray}
{\Gamma_{u_f d_i }^{H^\pm\,LR\,\rm{eff} } } &=& \sum\limits_{j = 1}^3
{\sin\beta\, V_{fj} \left( \frac{m_{d_i }}{v_d} \delta_{ji}-
  \epsilon^{ d}_{ji}\tan\beta \right), }
\\
{\Gamma_{u_f d_i }^{H^ \pm\,RL\,\rm{eff} } } &=& \sum\limits_{j = 1}^3
{\cos\beta\,  \left( \frac{m_{u_f }}{v_u} \delta_{jf}-
  \epsilon^{ u\star}_{jf}\tan\beta \right)V_{ji}}\,.\nonumber
 \label{Higgs-vertices-decoupling}
\end{eqnarray}
Thus, the Wilson coefficients $C_{L}^{qb}$ and $C_{R}^{qb}$ at the matching scale are given by
\begin{equation}
\renewcommand{\arraystretch}{1.5}
\begin{array}{l}
 C_{R(L)}^{qb} = \dfrac{{ -1}}{M_{H^{\pm}}^{2}} \; \Gamma_{qb}^{LR(RL),H^{\pm}} \; \dfrac{m_\tau}{v}\tan\beta  \,, 
 \end{array}
\end{equation}
with the vacuum expectation value $v\approx174{\rm GeV}$. Here we assumed that the Peccei-Quinn breaking for leptons is negligible, which means that the lepton-Higgs coupling are like in the 2HDM of type II. Note that for large Higgs masses and large values $\tan(\beta)$, the CP-odd and the heavy CP-even Higgs mass approach the charged one.

\begin{figure*}[t]
\centering
\includegraphics[width=0.3\textwidth]{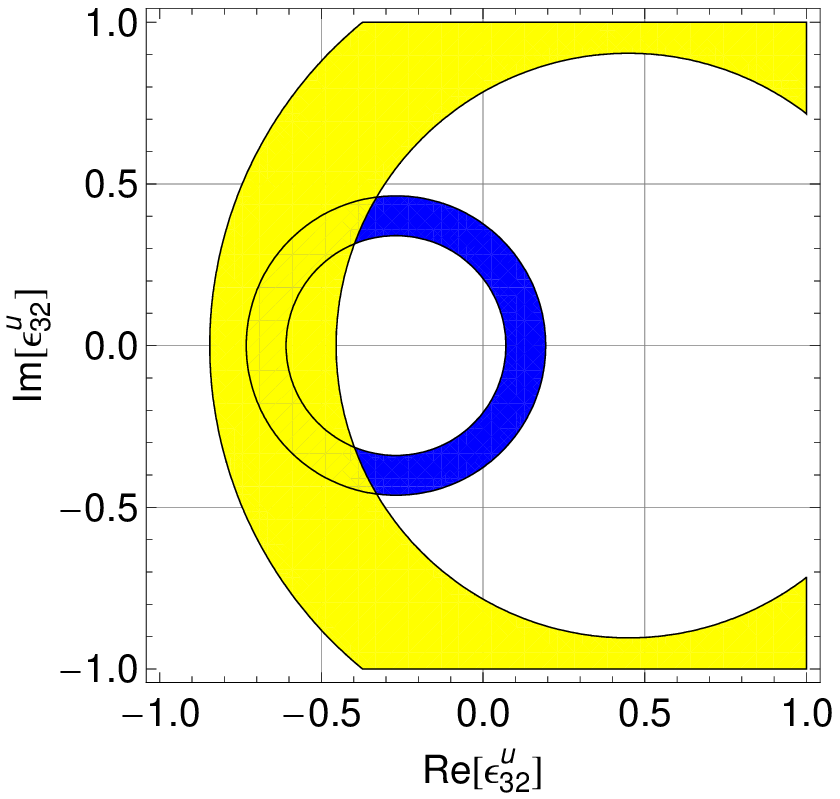}
\includegraphics[width=0.31\textwidth]{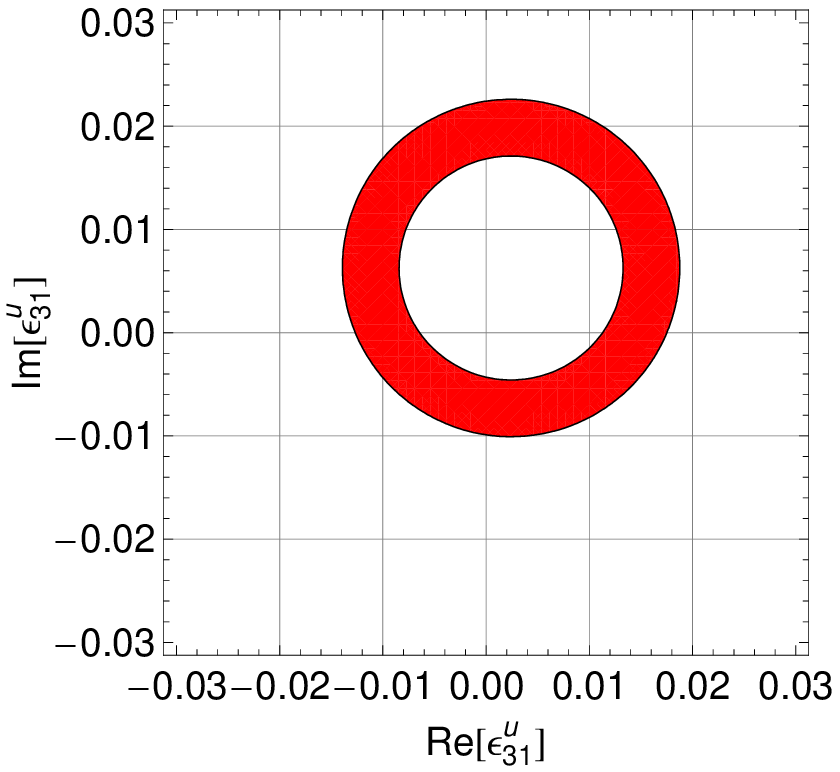}
\includegraphics[width=0.31\textwidth]{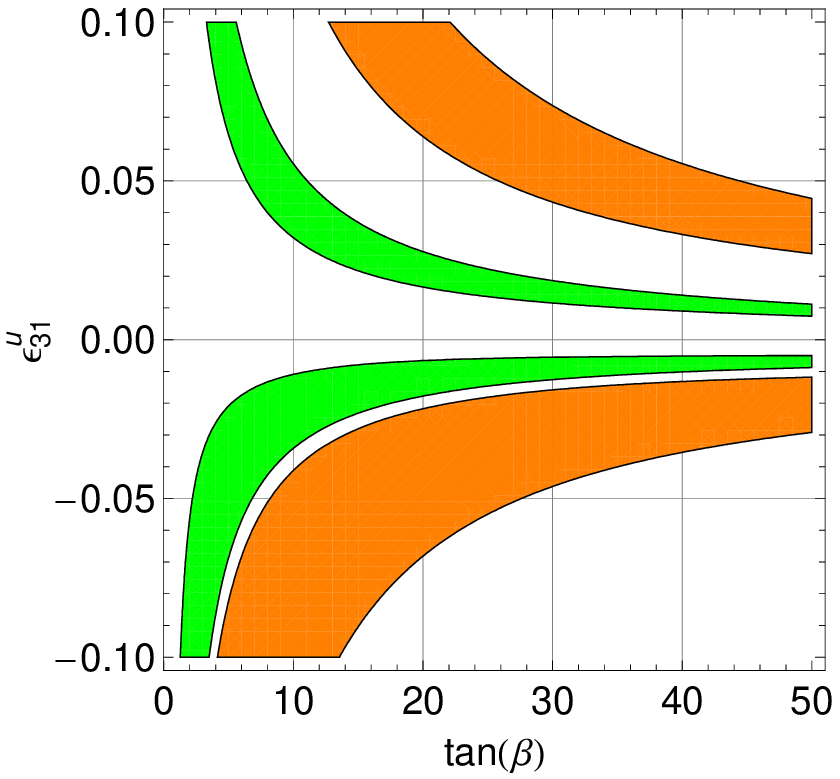}
\caption{Left: Allowed regions in the complex $\epsilon^u_{32}$-plane from \rd (blue) and \rds (yellow) for $\tan\beta=50$ and $m_H=500$~GeV. Middle:  Allowed regions in the complex $\epsilon^u_{31}$-plane from \btau. Right:  Allowed regions in the $\tan\beta$--$\epsilon^u_{31}$ plane from \btau for real values of $\epsilon^u_{31}$ and $m_H=400$~GeV (green), $m_H=800$~GeV (orange). The scaling of the allowed region for $\epsilon^u_{32}$ with $\tan\beta$ and $m_H$ is the same as for $\epsilon^u_{31}$. $\epsilon^u_{32}$ and $\epsilon^u_{31}$ are given at the matching scale $m_H$. \label{2HDMIII}}
\end{figure*}

\subsection{Experimental constraints}
First, note that all flavor-changing elements $\epsilon^d_{ij}$ are stringently constrained from FCNC processes in the down sector because of tree-level neutral Higgs exchange. Thus, they cannot have any significant impact on the decays we are interested in, and therefore we are left with~$\epsilon^d_{33}$.

Concerning the elements $\epsilon^u_{ij}$ we see that only $\epsilon^u_{31}$ ($\epsilon^u_{32}$) significantly effects \btau (\rd and \rds) without any CKM suppression. Furthermore, since flavor-changing top-to-up (or charm) transitions are not measured with sufficient accuracy, we can only constrain these elements from charged Higgs-induced FCNCs in the down sector. However, since in this case an up (charm) quark always propagates inside the loop, the contribution is suppressed by the small Yukawa couplings of the up-down-Higgs (charm-strange-Higgs) vertex involved in the corresponding diagrams. Thus, the constraints from FCNC processes are weak, and $\epsilon^u_{32,31}$ can be sizable. 

Of course, the lower bounds on the charged Higgs mass for a 2HDM of type II from $b\to s\gamma$ of 300~GeV \cite{Misiak:2006zs} must still be respected by our model, and also the results from direct searches at the LHC \cite{CMS} are in principle unchanged. Note that the recent CMS results even welcome a heavy Higgs ($H^0,A^0,H^\pm$) mass  around 500 GeV.

\subsection{\bdtau and \bdstau}

$\epsilon^d_{33}$ contributes to $C_R^{cb}$, and thus (as we see from Fig.~\ref{EFT}) cannot simultaneously explain \rd and \rds. Thus, we are left with $\epsilon^u_{32}$, which contributes to \bdtau and \bdstau via the Feynman diagram shown in Fig.~\ref{feynman-diagram}. In Fig.~\ref{2HDMIII} we see the allowed region in the complex $\epsilon^u_{32}$-plane, which gives the correct values for \rd and \rds within the $1\, \sigma$ uncertainties for $\tan\beta=50$ and $M_H=500$~GeV.

\subsection{\btau}

In principle, \btau can be explained either by using $\epsilon^d_{33}$ (as in 2HDMs with MFV) or by $\epsilon^u_{31}$, or by a combination of both (see right plot in Fig.~\ref{EFT}). However, $\epsilon^d_{33}$ alone cannot explain the deviation from the SM without fine tuning, while $\epsilon^u_{31}$ is capable of doing this. We see this from the right plot in Fig.~\ref{2HDMIII}, keeping in mind that $\epsilon^d_{33}$ generates $C_R^{ub}$, while $\epsilon^u_{31}$ generates $C_L^{ub}$.
\subsection{The quark mass matrix and fine tuning}
The naturalness criterion of 't Hooft states that the smallness of a quantity is only natural if a symmetry is gained in the limit in which this quantity is zero. This means, on the other hand, that large accidental cancellations, which are not enforced by a symmetry, are unnatural and thus not desirable. Let us apply this reasoning to the quark masses and CKM elements in the 2HDM. The quark mass matrices in the 2HDM of type III are given by
\begin{eqnarray}
m^{d(u)}_{ij}=v_{d(u)} Y^{d(u)}_{ij} + v_{u(d)} \epsilon^{d(u)}_{ij}\,.
\end{eqnarray}
Diagonalizing these quark mass matrices gives the physical quark masses and the CKM matrix. Using 't Hooft's naturalness criterion we can demand the absence of fine-tuned cancellations between $v_d Y^{d}_{ij}$ ($v_u Y^{u}_{ij}$) and $v_u \epsilon^d_{ij}$ ($v_d \epsilon^u_{ij}$). Thus, we require that the contributions of $v_u \epsilon^d_{ij}$ and $v_d \epsilon^u_{ij}$ to the quark masses and CKM matrix not exceed the physical measured quantities:
\begin{eqnarray}
|v_{u(d)} \epsilon^{d(u)}_{ij}|\leq \left|V_{ij}\right|\,{\rm max }\left[m_{d_i(u_i)},m_{d_j(u_j)}\right]\,.
\end{eqnarray}
From Fig.~\ref{2HDMIII}, we see that 't Hooft's naturalness criterion is satisfied if \rd, \rds and \btau are explained using $\epsilon^u_{32}$ and $\epsilon^u_{31}$, respectively. However, if \btau is explained using $\epsilon^d_{33}$, 't Hooft's naturalness criterion is violated either because the SM contribution to \btau is overcompensated or because $\left|v_u\epsilon^d_{33}\right|>m_b$. 

\section{Conclusions}

The decays \btau, \bdtau and \bdstau are an excellent probe of physics beyond the SM (complementary to the direct searches at the LHC), since they are sensitive to lepton flavor universality violating new physics, e.g., Higgs bosons. The BABAR Collaboration recently reported an excess both in \bdtau and \bdstau compared to the SM predictions \cite{BaBar:2012xj}. This evidence for new physics cannot be explained with a 2HDM of type II. Therefore, we proposed a 2HDM of type III with MSSM-like Higgs potential and flavor-violation in the up sector in order to explain these deviations from the SM. In fact, our model can account for the deviation of \rd and \rds from the SM predictions simultaneously and also bring \btau into agreement with experiment. This is even possible without significant fine tuning. Furthermore, all experimental constraints from other processes can be satisfied, and recent CMS results \cite{CMS} even welcome a mass around 500 GeV for the non-SM-like Higgs bosons of a 2HDM. In order to test the model, we propose to search for $A^0,H^0\to \overline{t}+c$ at the LHC.

\vspace{2mm}
{\it Acknowledgments.}--- 
{\small
We are grateful M. Procura for proofreading the article. A.C. thanks Eugenio Paoloni for useful discussions and bringing the $B\to D^{(*)}\tau \nu$ problem to his attention. This work is supported by the Swiss
National Science Foundation. The Albert Einstein Center for Fundamental Physics
is supported by the ''Innovations- und Kooperationsprojekt C-13'' of the
Schweizerische Universit\"atskonferenz SUK/CRUS.}

\bibliography{B-Dtaunu}

\end{document}